\documentclass[prl,twocolumn,preprintnumbers,superscriptaddress,groupedaddress,aps]{revtex4-2}
\usepackage[T1]{fontenc}
\usepackage[latin9]{inputenc}
\usepackage{amsmath}
\usepackage{graphicx}
\usepackage{color}
\usepackage{transparent}
\usepackage{bm}

\usepackage{mathrsfs}

\makeatletter

\providecommand{\tabularnewline}{\\}

\def \tr {\mathop{\rm tr}\nolimits}

\usepackage{babel}

\makeatother

\usepackage{babel}
\begin{document}
\title{Tracy-Widom distribution in four-dimensional super-Yang-Mills theories}
\author{Zoltan Bajnok$^{1}$, Bercel Boldis$^{1,2}$ and Gregory P. Korchemsky$^{3}$}
\affiliation{\textit{${}^1$HUN-REN
Wigner RCP, Konkoly-Thege Miklos ut 29-33, 1121 Budapest, Hungary}}
\affiliation{\textit{${}^2$Budapest University of Technology and Economics }
\textit{M\H{u}egyetem rkp. 3., 1111 Budapest, Hungary}}
\affiliation{\textit{${}^3$Institut de Physique Th\'eorique\footnote{Unit\'e Mixte de Recherche 3681 du CNRS}, Universit\'e Paris Saclay, CNRS,  91191 Gif-sur-Yvette, France}}
\begin{abstract}
Various observables in different four-dimensional superconformal Yang-Mills theories can be computed exactly
as Fredholm determinants of truncated Bessel operators. We exploit this relation to determine their dependence on the 't Hooft coupling constant. Unlike the weak coupling expansion, which has a finite radius of convergence,  the strong coupling expansion is factorially divergent, necessitating the inclusion of nonperturbative, exponentially small corrections. 
We develop a method to systematically compute these corrections and discuss the resurgent properties of the resulting transseries.
\end{abstract}

\preprint{IPhT-T24/008}

\maketitle

\section*{Introduction}

The Tracy-Widom distribution is a powerful tool to analyze a wide range of complex systems in physics. Discovered in the study of the statistics of the spacing of eigenvalues in random matrices, it was soon recognised to describe various phenomena in different fields, including quantum chaos, directed polymers, the Kardar-Parisi-Zhang equation, turbulence, etc. The universality of the Tracy-Widom distribution highlights the underlying connections between seemingly different physical phenomena~\cite{corwin2022harold}.

It was recently recognized  \cite{Beccaria:2022ypy} that the Tracy-Widom distribution describes 
a special class of observables in interacting four-dimensional superconformal $\mathcal N=2$ and $\mathcal N=4$ Yang-Mills theories (SYM) with the $SU(N_c)$ gauge group. 
They include the free energy on a sphere, vacuum expectation values of circular Wilson loops, correlation functions of infinitely heavy half-BPS operators, flux tube correlations.
A remarkable feature of these observables (denoted as ${\cal F}(g)$) is that, for an arbitrary 't Hooft  coupling constant $\lambda=g^2_{_{\rm YM}}N_c$ and to leading order in $1/N_c$, they can be expressed as determinants of certain semi-infinite matrices 
\begin{align}\label{eq:Flg}
e^{{\cal F} (g)}=\det_{1\leq n,m<\infty}(\delta_{nm}-K_{nm}(g )) \,.
\end{align}

The dependence on the coupling constant  $g=\sqrt\lambda/(4\pi)$ is encoded in the properties of the matrix $K_{nm}(g)$. Surprisingly, it exhibits a universal form across all observables
\begin{align} \label{eq:K_nm}
K_{nm}(g )=\int_{0}^{\infty}dx\,\psi_{n}(x)\chi\Bigl(\frac{\sqrt{x}}{2g}\Bigr)\psi_{m}(x)\,,
\end{align}
where the functions $\psi_{n}(x)$ satisfy the orthogonality condition $\int_{0}^{\infty}dx\psi_{n}(x)\psi_{m}(x)=\delta_{nm}$ and are given by the normalized Bessel functions 
\begin{equation}\label{psi}
\psi_{n}(x)=(-1)^{n}\sqrt{2n+\ell-1}J_{2n+\ell-1}(\sqrt{x})/\sqrt{x}\,.
\end{equation}
The parameter $\ell$ is a non-negative integer. 
 The function $\chi(\sqrt x/(2g))$ serves as a cutoff function in the integral (\ref{eq:K_nm}). It suppresses the contribution from large $x$, and is conventionally called the symbol of the matrix. Its explicit form and the corresponding value of $\ell$ depend on the choice of the observable.

The function \eqref{eq:Flg} is known in the mathematical literature as a Fredholm determinant of the truncated Bessel operator \cite{BasorEhrhardt03}. 
It depends on the symbol function $\chi(x)$ in a nontrivial way.  For $\chi(x)=\theta(1-x)$, the determinant (\ref{eq:Flg}) yields the Tracy-Widom distribution -- it describes the 
eigenvalue spacing at the hard edge of the Laguerre unitary ensemble \cite{Forrester:1993vtx}. It can be computed explicitly in terms of solutions to the Painlev\'e~V equation \cite{Tracy:1993xj}.

In application to the four-dimensional SYM theories, we encounter the symbol functions of different forms:
In planar $\mathcal N=4$ SYM, the expectation value of the circular Wilson loop is described by  (\ref{eq:Flg}) and (\ref{eq:K_nm}) for $\chi_{_{\rm W}}(x)=-(2\pi)^2/x^2$ and $\ell=2$~\cite{Beccaria:2023kbl}.
The flux tube correlations correspond to $\chi_{\text{f.t.}}(x)=2/(1-e^{x})$ and $\ell=0,1$ \cite{Beisert:2006ez,Belitsky:2019fan,Basso:2020xts}. 
The four-point correlation function of infinitely-heavy half-BPS operators depends on two cross ratios ($y$ and $\xi$), and it equals  the square of the octagon form factor \cite{Coronado:2018ypq,Coronado:2018cxj}. This form factor is given by  (\ref{eq:Flg}) and (\ref{eq:K_nm}) for $\chi_{\rm oct}(x)=(\cosh y+\cosh\xi)/(\cosh y+\cosh\sqrt{x^{2}+\xi^{2}})$ and nonnegative integer $\ell$  \cite{Kostov:2019stn,Kostov:2019auq,Belitsky:2020qrm,Belitsky:2020qir}. In $\mathcal N=2$ SYM, the free energy on the sphere and the circular Wilson loop can be computed using localization \cite{Pestun_2017}. The leading nonplanar corrections to both quantities are given by (\ref{eq:Flg}) and (\ref{eq:K_nm})  for $\chi_{\text{loc}}(x)=-1/\sinh^{2}(x/2)$ and $\ell=1,2$ \cite{Beccaria:2020hgy,Beccaria:2023kbl}.

The presented examples only hint at the broader applicability of (\ref{eq:Flg}) and (\ref{eq:K_nm}).  We anticipate that the function $\mathcal F(g)$ when evaluated for various symbol functions $\chi(x)$, will enter the  calculation of other important observables in four-dimensional gauge theories.

In this letter, we describe a systematic approach to finding the dependence of the function \eqref{eq:Flg} on the coupling constant for a sufficiently smooth symbol function $\chi(x)$.

\smallskip
\paragraph*{\textbf{Weak coupling expansion.}}

At weak coupling, changing the integration variable in \eqref{eq:K_nm} as $x\to (2g)^2 x$ and expanding the integrand at small $g$, one obtains $K_{nm}(g) = O((g^2)^{n+m+\ell-1})$. This enables us to expand the determinant \eqref{eq:Flg} in powers of the  matrix
\begin{align}\label{eq:Flgweak}
\mathcal F(g) = -\tr K-\frac{1}{2}\tr(K ^2)-\frac{1}{3}\tr(K ^3)+\dots 
\end{align}
where ${\rm tr}(K ^L) = O(g^{2L(\ell+1)})$. The leading term in \eqref{eq:Flgweak} is given by \cite{Belitsky:2020qir}
\begin{align}\label{trB}
\tr K = g^{2(\ell+1)}\sum_{k=0}^\infty (-g^2)^{k}{q_{\ell+k+1} (2\ell+2k)!\over k! (2\ell+k)! (\ell+k+1)!^2} \,,
\end{align}
where $q_k=2k \int_0^\infty dx\, x^{2k-1} \chi(x)$ is a moment of the symbol function. 
For this integral to be well-defined, the function $\chi(x)$ has to vanish sufficiently fast at infinity. One can verify that all the previously defined symbols, except $\chi_{_{\rm W}}(x)$, exhibit the behavior $\chi(x)\sim c\,e^{-x}$, where the normalization constant is $c_{\text{f.t.}}=-2$,  $c_{\text{oct}}= 2(\cosh y + \cosh \xi)$ and $c_{\text{loc}}=-4$.  

For  $\chi_{_{\rm W}}(x)=-(2\pi)^2/x^2$, it follows from \eqref{eq:K_nm} that $K_{nm}(g)\sim  g^2$. As a result, the relation \eqref{trB} is replaced with $\tr K= -a\, (\pi g)^2$,
where $a=16\sum_{n\ge 1} \int_0^\infty dx\,\psi_n^2(x)/x=4/\ell$. In this case, the weak coupling expansion \eqref{eq:Flgweak} can be resummed to all orders leading to the exact result \cite{Beccaria:2023kbl}
\begin{align}\label{F-ex}
\mathcal F_{_{\rm W}}(g) = \log \Big(\Gamma(\ell)(2\pi g)^{1-\ell} {\tt I}_{\ell-1}(4\pi g)\Big)\,,
\end{align}
where $\tt I_{\ell-1}$ is a modified Bessel function of the first kind. The relation \eqref{F-ex} holds for any coupling constant. 

As follows from \eqref{F-ex}, $e^{\mathcal F_{_{\rm W}}(g)}$ is an entire function of the coupling constant.
 For the remaining symbols satisfying $\chi(x)\sim c\,e^{-x}$ as $x\to\infty$, the weak coupling expansion \eqref{eq:Flgweak} exhibits a finite radius of convergence. 
 Indeed, replacing the moments  in \eqref{trB} with their large $k$ behaviour, $q_k\sim c \,(2k)!$, we obtain 
\begin{align}\label{trB-log}
\tr K  \sim  (-1)^\ell \frac{c}{4\pi} \log \left(g^2-g_\star^2\right)\,,
\end{align}
where $g_\star^2=-1/16$, or equivalently $\lambda_\star=-\pi^2$.
Repeating the same analysis for $\tr(K^L)$, one finds that each term of the expansion \eqref{eq:Flgweak} has a logarithmic singularity at $g^2=g_\star^2$, leading to
\begin{align}
\mathcal F(g)\sim \kappa \log \left(g^2-g_\star^2\right),
\end{align}
where $\kappa = \arcsin(c/2) \left(\arcsin(c/2)-(-1)^{\ell} \pi \right)/(2\pi^2)$.
 
\smallskip 
\paragraph*{\textbf{Strong coupling expansion.}} 

The strong coupling expansion of \eqref{eq:Flg} takes the form \cite{Beccaria:2022ypy}
\begin{equation}\label{F-gen}
{\cal F}(g)=-gA_{0}+\frac{1}{2}A_{1}^{2}\log g+B+f(g)+\Delta f(g)\,,
\end{equation}
where the coefficients depend on the symbol function and parameter $\ell$.
The first three terms on the right-hand side of \eqref{F-gen} follow from the Szeg\H{o}-Akhiezer-Kac formula for the truncated Bessel operator \cite{Belitsky:2020qir}. 
The two remaining terms in \eqref{F-gen} describe corrections vanishing at large $g$. 

The function $f(g)$ is given by an asymptotic series
\begin{equation}\label{f}
f(g)=\sum_{k=1}^{\infty}\frac{A_{k+1}}{2k(k+1)}g^{-k}\,.
\end{equation}
The expansion coefficients can be found to any order using the method of differential equations \cite{doi:10.1142/S0217979290000504,Korepin:1993kvr,Tracy:1993xj}.
The first few coefficients look as \cite{Belitsky:2020qrm,Belitsky:2020qir}
\begin{align}\notag\label{As}
 & A_{0}=2I_{0}\,,\qquad A_{1}=\sqrt{\ell_{\beta}^{2}-\ell^{2}}\,,
 \\\notag
 {}& A_{2}=-\frac{(4\ell_{\beta}^{2}-1)}{4}I_{1}\,,\qquad  A_{3}=-\frac{3(4\ell_{\beta}^{2}-1)}{16}I_{1}^{2}\,,
 \\
 {}&
A_{4}=-\frac{(4\ell_{\beta}^{2}-1)(16I_{1}^3+(4\ell_{\beta}^{2}-9)I_{2})}{128}\,,\quad \dots
 && 
\end{align}
Here the notation  $\ell_{\beta}=\ell+\beta$  was introduced,  where $\beta$ defines the behaviour of the symbol function at the origin, $1-\chi(x) \sim b x^{2\beta}$ as $x\to 0$. For the symbol functions defined above we have $\beta_{\text{loc}} = -\beta_{\text{oct}} = 2\beta_{\text{f.t.}} =-1$. The expression for the coefficient $B$ in \eqref{F-gen} is more involved and it can be found in \cite{Belitsky:2020qir}. 

The functions $I_n=I_n(\chi)$ are defined as  
\begin{align}\label{prof}
I_{n}(\chi)=\int_{0}^{\infty}\frac{dx}{\pi}\frac{(x^{-1}\partial_{x})^{n}}{(2n-1)!!} x\partial_{x}\log(1-\chi(x))
\end{align}
where $n\ge 0$. For the symbols under consideration they can be expressed in terms of odd Riemann zeta values
\begin{align}\label{Is}
I_n^\text{loc}=-I_n^\text{oct}=2I_n^\text{f.t.} = (-1)^{n-1}(1-2^{2-2n}) {2 \zeta_{2n-1}\over \pi^{2n-1}}
\end{align}
Here the octagon function $I_n^\text{oct}$ is evaluated for $\xi=y=0$. We will use this specific case throughout our analysis. 
In general, for any values of $\xi$ and $y$, the function $I_n^\text{oct}$ can be expressed in terms of ladder integrals \cite{Belitsky:2020qir}.

Notice that all coefficients \eqref{As}, except $A_1$, depend on the parameters $\ell$ and $\beta$ only through their sum $\ell_{\beta}$. For $\ell_{\text{loc}} =\ell_{\text{oct}}+2$ we have $\ell_{\beta}^\text{loc}=\ell_{\beta}^\text{oct}$ and, in virtue of \eqref{Is}, the coefficients satisfy $A_k^\text{loc}=(-1)^{k+1}A_k^\text{oct}$ for $k\ge 1$. Together with \eqref{f} this leads to \cite{Beccaria:2022ypy}
\begin{align}\label{for}
f_{\text{loc}}(g) = f_{\text{oct}}(-g)
\end{align} 
This relation is formal because the expansion coefficients $A_{k}$ grow factorially at large $k$. As a consequence, the series \eqref{f} is only asymptotic and it is plagued with Borel singularities. 

For the strong coupling expansion \eqref{F-gen} to be well defined, it has to be supplemented with the nonperturbative, exponentially small corrections $\Delta f(g)$. For the symbol functions introduced above, they take the form of the transseries \cite{Basso:2009gh,Beccaria:2022ypy}
\begin{align}\label{trans}
 \Delta f(g) = \sum_{n\ge 1} \left(g^a e^{-8 \pi g x_1} \right)^n \left[ A^{(n)}_{1}+\sum_{k=1}^{\infty}\frac{A^{(n)}_{k+1}}{2k(k+1)}g^{-k}\right]
\end{align}
where $a=0$ for the flux tube and $a=1$ in the other cases, while
  $x_1$ is a solution to $\chi(2i\pi x_1)=1$ closest to the origin (see \eqref{symb} and \eqref{Phi} below). 
Our next step is to develop a systematic method to compute the expansion coefficients $A^{(n)}_{k+1}$ and, then, investigate the resurgent properties of the series  \eqref{trans}. 
 
\section*{Nonperturbative corrections} 
 
\paragraph*{\textbf{Truncated Bessel operator.}} 

It is advantageous to represent the semi-infinite matrix \eqref{eq:K_nm} as defining the matrix elements $K_{nm} = \langle \psi_n|\bm K_\chi|\psi_m \rangle$ of an integral operator
\begin{align}\label{Bes}
\bm K_\chi  \phi(x) = \int_{0}^{\infty}dy\, K(x,y)\chi\Big( {\sqrt{y}\over 2g}\Big)\phi(y) 
\end{align}
where $\phi(x)$ is a test function and $K(x,y)$ is given by an infinite sum of normalized Bessel functions \eqref{psi}, $K(x,y)=\sum_{n\geq1}\psi_{n}(x)\psi_{n}(y)$.

The determinant of the semi-infinite matrix in \eqref{eq:Flg} coincides with a Fredholm determinant of the
operator \eqref{Bes} 
\begin{align}
\mathcal F(g) = \log \det(1- \bm K_\chi)
\end{align}
Using the method of differential equations \cite{doi:10.1142/S0217979290000504,Korepin:1993kvr,Tracy:1993xj},  one can show that the function $\mathcal F(g)$ satisfies the equation \cite{Belitsky:2019fan,Belitsky:2020qrm,Belitsky:2020qir}
\begin{equation}\label{dF}
 \partial_{g}{\cal F}(g)=- \frac12\int_{0}^{\infty}dz\,  z \partial_{z}\chi(z) q^2(z,g) \partial_z \partial_g \log q(z,g)
\end{equation}
where $q(z,g)$ is an auxiliary function. It is defined as a matrix element of the resolvent of the truncated Bessel operator
\eqref{Bes}
\begin{align}\label{q}
q(z,g) = \langle x\vert\frac{1}{1-\bm{K}_{\chi}}\vert\phi_{0}\rangle\,,\qquad z={\sqrt x/(2g)},
\end{align}
where $\phi_0(x) = J_\ell(\sqrt x)$ is a reference state. 
 
The function \eqref{q} has the following properties. It is an entire function of $z$ for any $g$ and has a parity $q(-z,g)=(-1)^\ell q(z,g)$. In addition, it satisfies an infinite system of integral equations
\begin{equation}\label{q-int}
\int_{0}^{\infty}dx \,(1-\chi(x)) J_{2n+\ell-1}(2gx)q(x,g) =0
\end{equation}
where $n\ge 1$, as well as a differential equation \cite{Belitsky:2020qrm,Belitsky:2020qir}
\begin{equation}\label{eq:diffeq}
\Big((g\partial_{g})^{2} + 4g^{2}x^{2}-\ell^{2}+2g^2\partial^2_{g}\mathcal F(g)\Big) q(x,g)=0 
\end{equation}
Notice that this equation involves $\mathcal F(g)$ which depends on $q(x,g)$ in a nontrivial way (see \eqref{dF}). 

By combining relations \eqref{dF}, \eqref{q-int}, and \eqref{eq:diffeq}, we can compute the derivative $\partial_g \mathcal F(g)$ for any coupling constant $g$.

\smallskip
\paragraph{\textbf{Asymptotic solution.}}

To solve the equations \eqref{q-int} and \eqref{eq:diffeq} at strong coupling, we  perform a Wiener-Hopf type decomposition of the symbol function
\begin{align}\label{symb}
1-\chi(x)=bx^{2\beta}\Phi(x)\Phi(-x) 
\end{align}
where $\beta$ is (half) integer. The function $\Phi(x)$ is normalized as $\Phi(0)=1$ and it takes the form
\begin{align}\label{Phi}
\Phi(x)=\prod_{n=1}\frac{1-\frac{ix}{2\pi x_{n}}}{1-\frac{ix}{2\pi y_{n}}}
\end{align}
It is analytic in the upper half plane and has an infinite set of poles and zeros located at $x=-2i\pi x_n$ and $x=-2i\pi y_n$, respectively. The symbol function \eqref{symb} is represented by the parameters $b,\beta, x_n, y_n$ (with $n\ge 1$).

The symbol functions introduced above have the form \eqref{symb}, e.g.
\begin{align}\label{cases}
 \Phi_{\rm loc}(x) ={1\over \Phi_{\rm oct}(x)} =[\Phi_{\text{f.t.}}(x)]^2 =  \pi \left[{\Gamma(1-{ix\over 2\pi})\over\Gamma({1\over 2}-{ix\over 2\pi})}\right]^2
\end{align}
Notice that the poles and zeros of the first two functions are double degenerate. Moreover,  the poles of one coincide with the zeros of the other, and vice versa, e.g.
$x_n^{\rm oct}=y_n^{\rm loc}=n$ and $y_n^{\rm oct}=x_n^{\rm loc}=n-\frac12$.~\footnote{The quoted values of $x_n^{\rm oct}$ and $y_n^{\rm oct}$ correspond to the kinematic point $y=\xi=0$ of the octagon. In general, $x_n^{\rm oct}$ and $y_n^{\rm oct}$ are non-trivial functions of $y$ and $\xi$.}

Equations \eqref{q-int} and \eqref{eq:diffeq} play complementary roles in determining the function $q(x,g)$: the former can be used to derive the correct ansatz for $q(x,g)$ at strong coupling and the latter  serves the purpose of fixing all the remaining unknown parameters within the chosen ansatz.

The integral equation \eqref{q-int} can be solved at strong coupling using an approach presented in \cite{Beccaria:2022ypy}.
The resulting expression for $q(x,g)$ looks as
\begin{align}\label{q-a}
q(x,g) = {e^{2igx}\over \Phi(-x)} h(x,g) + (-1)^\ell {e^{-2igx}\over \Phi(x)} h(-x,g) 
\end{align}
The coefficient function $h(x,g)$ is given by a transseries running in powers of $e^{-8\pi gx_n}$, where the parameters $x_n$ are defined in \eqref{Phi}. 
For the cases in \eqref{cases}, $x_{n}$ becomes an integer
multiple of $x_{1}$, leading to the strong coupling expansion of $h(x,g)$ being
\begin{equation}\label{f-trans}
h(x,g)=\sum_{n\ge 0}e^{-8\pi g x_1 n}  \sum_{k\ge 0} h_{k}^{(n)}(x) g^{-k-1/2}
\end{equation}

The requirement for $q(x,g)$ to be an entire function of $x$ imposes an additional constraint on the function $h(x,g)$. It follows from the condition that the product $(1-\chi(x))q(x,g)$ has to vanish for $x\to -2\pi i x_n$ (see \eqref{symb} and \eqref{Phi}). 
Depending on the explicit form of $x_n$ in \eqref{cases}, we have to distinguish two cases.
For $x^{\rm oct}_n=n x_1$, we replace $q(x,g)$ with \eqref{q-a} and \eqref{f-trans} to get from the above condition (for $n\ge 1$)
\begin{align}\label{qc} 
\lim_{x\to-2i\pi n x_{1}}{}&  h^{(n)}_{k}(x)\Phi(x)=(-1)^{\ell+1}
 h_k^{(0)}(2i\pi n x_{1})\Phi(2i\pi n x_{1})
\end{align}
For $x^{\rm loc}_m=(2m-1)x_1$, the relation \eqref{qc} holds for odd $n=2m-1$. For even $n=2m$, the right-hand side of \eqref{qc} should vanish in order to avoid the appearance of unphysical poles of $q(x,g)$ at $x=-4i\pi m x_1$.

The left-hand side of \eqref{qc} is different from zero only if $h^{(n)}_{k}(x)$ has a pole at $x=-2i\pi n x_{1}$.
Equation \eqref{qc} expresses the residues of $h^{(n)}_{k}(x)$ at this pole
in terms of the `perturbative' coefficient function $h_k^{(0)}(2i\pi n x_{1})$.  
Substituting \eqref{q-a} and \eqref{f-trans} into \eqref{eq:diffeq} yields additional relations for the coefficient functions $h_k^{(n)}(x)$. Solving these relations allows us to determine the coefficient functions $h_k^{(n)}(x)$ and subsequently compute the expansion coefficients in the transseries \eqref{trans}. 

As a first step, we plug \eqref{F-gen} and \eqref{q-a} in \eqref{eq:diffeq} and equate to zero the coefficients of  $1/g^k$ and $e^{-8 \pi g x_1 n}$. For $n=0$, this gives $h_k^{(0)}(x)$ in terms of the coefficients $A_m^{(0)}\equiv A_m$ (with $m\le k$) defined in \eqref{f} and \eqref{As}. For $n=1$, we find that $h_k^{(1)}(x)$ has poles at $x=-2\pi ix_1$. Their residues depend on the coefficients $A_m^{(1)}$ (with $m\le k$). Substituting the resulting expressions for $h_k^{(0)}(x)$ and $h_k^{(1)}(x)$ into \eqref{qc} we obtain the relations between the $A^{(1)}-$ and $A^{(0)}-$coefficients. Following the same procedure for  $n\ge 2$, we can iteratively compute all coefficients $A_m^{(n)}$ in terms of the leading coefficients $A_m$ defined in \eqref{As}. 

To carry out the above procedure, we have to specify the symbol function \eqref{symb} and \eqref{Phi}.

\smallskip
\paragraph{\textbf{Results.}}

In the case of the flux tube, for $\ell=0$ and $\ell_\beta=-1/2$, we find from \eqref{As} that $A_0=-\pi/2$, $A_1^2=1/4$ and $A_n=0$ for $n\ge 2$. According to \eqref{cases}, the function $\Phi_{\text{f.t.}}(x)$ has zeros at $x_1=1/2$ and $x_n=(2n-1)x_1$. Going through the calculation we determined the coefficients,
$A_{1}^{(n)}= {\left(1-3 (-1)^n\right) }/{(8 n)}$ and
 $A_{k\geq 2}^{(n)}=0$, and finally obtained from \eqref{F-gen} and \eqref{trans}
\begin{align}\notag
{\cal F}_{\text{f.t.}}(g)
{}&= {g\pi\over 2}+\frac18\log \Big({g\pi\over 2}\Big) +\sum_{n\ge 1} A_{1}^{(n)} e^{-4\pi g n}
\\
{}&=\frac{3}{8}\log\cosh(2\pi g)-\frac{1}{8}\log\frac{\sinh(2\pi g)}{2\pi g}
\end{align} 
in agreement with the known exact result \cite{Beccaria:2022ypy}. Similar formulas can be obtained for higher $\ell$'s too.   

For the two remaining symbols, the `perturbative' part of \eqref{F-gen}  is given by the asymptotic series,
$f_{\text{oct}}(g)$ and $f_{\text{loc}}(g)$ depending on $\ell_{\text{oct}}$ and $\ell_{\text{loc}}$, respectively. For $\ell_{\text{loc}} =\ell_{\text{oct}}+2$ they are related to each other by the sign change of the coupling constant (see \eqref{for}). 
The calculation of the nonperturbative correction \eqref{trans} becomes more intricate compared to the previous case because the zeros $x_n^{\rm oct}$ and $x_n^{\rm loc}$ are double degenerate. This can be handled by slightly splitting them a bit away, performing the calculations and taking the coinciding limit carefully. 

To save space, we present only the first few terms of the transseries \eqref{trans} evaluated for $\ell_{\text{loc}}=2$ and $\ell_{\text{oct}}=0$. 
For the octagon we find~\footnote{Higher order terms of the transseries \eqref{Delta-oct} and \eqref{Delta-loc}  can be found in a Mathematica notebook attached to the submission.} 
\begin{align}\notag\label{Delta-oct}
{}& \Delta f_{\text{oct}}(g)  = \frac{i \pi g'}{4}e^{-8\pi g}\bigg[1-\frac{7}{4( 4\pi g') }-\frac{63}{32(4\pi g')^{2}}\bigg]   
\\
{}&  +\frac{(\pi g')^{2}}{32}e^{-16\pi g}\bigg[1+\frac{\frac{81 i}{4}-\frac{7}{2}}{4 \pi g' }+\frac{-\frac{1431 i}{32}-\frac{3}{4}}{(4\pi g')^{2}}\bigg] +... 
\end{align}
where we changed the expansion parameter to $g'=g+\log (2)/\pi$ to avoid the appearance of $\log(2)$. Similarly, 
\begin{align}\notag\label{Delta-loc}
{}&\Delta f_{\text{loc}}(g)=2i \pi g''  e^{-4\pi g}  \bigg[1+\frac{8 \log 2+\frac{1}{2}}{4\pi g''}+\frac{12 \log
   2-\frac{15}{8}}{(4\pi g'')^2}\bigg]  
\\
{}&+2(\pi g'')^2 e^{-8\pi g}  \bigg[1+\frac{16 \log 2+1}{4\pi g''}+\frac{64\log ^2 2+32 \log 2-3}{(4\pi g'')^2}\bigg] +... 
\end{align}
where $g''=g-\log(2)/\pi$. The two couplings are related as $g'\to -g''$ for $g\to -g$.
The transseries \eqref{Delta-oct} and \eqref{Delta-loc} have different form and do not satisfy the relation \eqref{for}. 

A close examination shows that the series 
in \eqref{Delta-oct} and \eqref{Delta-loc} suffer from Borel singularities. 
Since the sum of functions $f(g)+\Delta f(g)$ in \eqref{F-gen} must be unambiguous, there exist highly non-trivial resurgence relations between the perturbative series \eqref{f} and those appearing in \eqref{Delta-oct} and \eqref{Delta-loc}.  
 
\renewcommand{\arraystretch}{1.75}
\begin{table*}[t]
\begin{centering}
\medskip
\begin{tabular}{|l|l|l|}
\hline 
$a^{_{(1)}}_{0}=-\frac{1}{32\pi}$ & $a^{_{(1)}}_{1}/a^{_{(1)}}_{0}=8\log2+\frac{1}{2}$ & $a^{_{(1)}}_{2}/a^{_{(1)}}_{0}=12\log2-\frac{15}{8}$\tabularnewline
\hline 
$a^{_{(2)}}_{0}= -{i\over 1024\pi}$ & $a^{_{(2)}}_{1}/a^{_{(2)}}_{0}=16\log2+1 $ & $a^{_{(2)}}_{2}/a^{_{(2)}}_{0}= 64(\log 2 )^2+32 \log 2-3 $ \tabularnewline
\hline 
$b^{_{(1)}}_{0}=-\frac{16}{\pi}$ & $b^{_{(1)}}_{1}/b^{_{(1)}}_{0}=-\frac{7}{4}$ & $b^{_{(1)}}_{2}/b^{_{(1)}}_{0}=-\frac{63}{32}$\tabularnewline
\hline 
$b^{_{(2)}}_{0}=-{256 i\over\pi}$ & $b^{_{(2)}}_{1}/b^{_{(2)}}_{0}=-\frac{7}{2}$ & $b^{_{(2)}}_{2}/b^{_{(2)}}_{0}=-\frac{3}{4}$\tabularnewline
\hline 
$c^{_{(2)}}_{0}=0$ & $c^{_{(2)}}_{1}/b^{_{(2)}}_{0}=\frac{81i}{4}$ & $c^{_{(2)}}_{2}/b^{_{(2)}}_{0}=-\frac{1431i}{32}$\tabularnewline
\hline 
\end{tabular}
\par\end{centering}
\caption{The coefficients of the strong coupling expansion extracted from the generalized Borel transform.}
\label{abcd}
\end{table*}  

\section*{Resurgence relations}

To investigate the resurgence properties of the strong coupling expansion \eqref{F-gen}, we performed a high precision calculation of the first $400$
terms of the perturbative series $f_{\text{oct}}(g)$. It proves convenient to change the expansion parameter to $g'=g+\log(2)/\pi$. 

Analysis of the numerical data revealed a factorial behavior in the expansion coefficients 
\begin{align}\label{alpha}
f_{\text{oct}}(g)=\sum_{n\ge 1}{\alpha_n \over (4\pi g')^n}\,,\qquad\quad 
\alpha_{n} \sim  \Gamma_{n+1} \,,
\end{align}
where $\Gamma_{n+1}$ is the shorthand notation for $\Gamma(n+1)$.
We therefore expect that a generalized Borel transformation 
\begin{equation}\label{Bk}
\mathcal B_{\delta}(s)=\sum_{n=0}^{\infty}\alpha_{n}\frac{ s^{n+\delta}}{\Gamma_{n+\delta+1}} 
\end{equation}
should have singularity for real $s$.  Indeed, we used the first $400$ terms in \eqref{Bk} to construct the diagonal Pade approximant
for $\mathcal B_{\delta=0}(s)$. We observed that its poles condense on the real axis for $s<-1$ and $s>2$, indicating that 
the Borel transform has cuts along the real axis that start at $s=-1$ and $s=2$, and potentially extend to higher values as well. 

Based on numerical studies, we can parameterize the large order behaviour of the coefficients \eqref{alpha} as 
\begin{align}\notag\label{a-gen}
{}& \alpha_n = \sum_{k\ge 0} (-1)^n \bigg( a_k^{(1)} \Gamma_{n+1-k} +a_k^{(2)}  {\Gamma_{n+2-k} \over 2^{n+2-k}} +\dots \bigg)
\\ 
{}&
+\sum_{k\ge 0} \bigg(b_k^{(1)}  {\Gamma_{n+1-k} \over 2^{n+1-k}} +b_k^{(2)}  {\Gamma_{n+2-k} \over 4^{n+2-k}} 
+\dots \bigg).
\end{align}
The terms on the first and second lines are of the form $(-1)^n a_k^{(p)}\Gamma_{n+p-k}/p^{n+p-k}$ and $b_k^{(p)}\Gamma_{n+p-k}/(2p)^{n+p-k}$, respectively,
with $p$ positive integers. They produce cuts of the Borel transform \eqref{Bk} at $s=-1,-2, \dots$ and
$s=2,4,\dots$ related to the locations of the non-perturbative corrections \eqref{trans}. These are logarithmic cuts for $\delta=0$ and square root cuts for half-integer $\delta$.

Choosing $\delta=\frac{1}{2}$ we can desingularise the square root cut of $\mathcal B_{\frac12}(s)$ at $s=-1$ by a coordinate change in the Borel plane \cite{Aniceto:2018uik,Costin:2020pcj}.
This allowed us to extract the coefficients $a_{k}^{_{(1)}}$
for $k=0,1,\dots$ with $65,61,\dots$ digit
precisions (see Table \ref{abcd}).
 Similarly, we extracted the coefficients $b_k^{_{(1)}}$  with $42,40,\dots$ digits precisions  from the behaviour of $\mathcal B_{\frac12}(s)$ around $s=2$. We found that these coefficients grow factorially at large $k$
\begin{align}
b_{k}^{(1)}= -{2\over\pi}\left(c_{0}^{(2)}{\Gamma_{k+1}\over 2^{k+1}}+c_{1}^{(2)}{\Gamma_{k}\over 2^{k}}+c_{2}^{(2)}{\Gamma_{k-1}\over 2^{k-1}}+...\right) 
\end{align}
We extracted the coefficients $a_k^{(2)}$ for $k=0,1,2,\dots$ with $23,20,17,\dots$ digits precision from the behaviour of $\mathcal B_{\frac32}(s)$ around $s=-2$ and the  $b_k^{(2)},c_k^{(2)}$ coefficients with  $8,6,5,\dots$ digits form the behaviour around $s=4$. 

We verify that, in a complete agreement with the resurgence relations (see reviews \cite{Aniceto:2018bis,Dorigoni:2014hea} and references therein), the ratio of the coefficients $b_k^{(1)}/b_0^{(1)}$ and $(b_k^{(2)}+c_k^{(2)})/b_0^{(2)}$ coincide with the coefficients of $1/(4\pi g')^k$
in the two series on the first and second line of \eqref{Delta-oct}, respectively. 
This ensures that the ambiguities generated by Borel singularities cancel in the sum $f_{\text{oct}}(g)+\Delta f_{\text{oct}}(g)$.  In particular, the imaginary part arising from integrating the Borel transform slightly above the real axis,
\begin{align}
f_{\text{oct}}(g) = 4\pi g'\int_0^{\infty e^{i\varepsilon}}\! ds\, \mathcal B_{0}(s) \, e^{-4\pi g' s},
\end{align}
cancels against the imaginary part coming from the nonperturbative function $\Delta f_{\text{oct}}(g)$. Consequently, $f_{\text{oct}}(g)+\Delta f_{\text{oct}}(g)$ approximates a real function of $g$.  
 
We also observe that the ratio of the coefficients $a_k^{(1)}/a_0^{(1)}$ and $a_k^{(2)}/a_0^{(2)}$ (see Table \ref{abcd}) coincide with the coefficients of $1/(4\pi g')^k$
in the two series on the first and second lines of \eqref{Delta-loc}, respectively. This property is an immediate consequence of the relation \eqref{for} which implies that the Borel transforms of $f_{\text{loc}}(g)$ and $f_{\text{oct}}(g)$ are related to each other as $\mathcal B^{\text{loc}}_{0}(s) = \mathcal B^{\text{oct}}_{0}(-s)$. Thus, the discontinuity of $\mathcal B_{0}^{\text{oct}}(s)$ across the cuts at negative $s$ has to match nonperturbative corrections to the function  $\Delta f_{\text{loc}}(g)$.

This suggests that, for $\ell_{\text{loc}} =\ell_{\text{oct}}+2$, the  transseries $f_{\text{oct}}(g)+\Delta f_{\text{oct}}(g)$ and  $f_{\text{loc}}(g)+\Delta f_{\text{loc}}(g)$, define the asymptotic expansion of the \textit{same} function, $F(g)$ and $F(-g)$, respectively, across the Stokes line at ${\rm Re}(g)=0$.

We verified that our analytical results for ${\cal F}_{\text{oct}}(g)$ and ${\cal F}_{\text{loc}}(g)$ (see \eqref{F-gen}) agree with high-precision numerical evaluation of  
the determinant \eqref{eq:Flg} for various values of $g$. By combining the weak and strong coupling expansions, we can effectively determine the two functions for any given value of the 't Hooft coupling.

Finding non-perturbative corrections at strong coupling is a notoriously difficult problem.  Currently, there are very few examples of resurgence in $ {\cal N}=4$ SYM \cite{Basso:2009gh,Dorigoni:2015dha,Aniceto:2015rua,Arutyunov:2016etw,Dorigoni:2021guq}. Our results provide a systematic treatment and new insights into the strong coupling regime of four-dimensional superconformal gauge theories, inviting further investigations within the AdS/CFT correspondence. In a dual string theory description, the obtained non-perturbative corrections might arise from world-sheet instantons describing specific world-sheet configurations that are intricately linked to the observables under study. In addition, the technique outlined above has important applications for determining the asymptotic behaviour of determinants of Bessel operators (see e.g. \cite{BasorEhrhardt03,BasorChen03}). 
 
 \section*{Acknowledgments}

We thank Benjamin Basso and Ivan Kostov for useful discussions and Annamaria Sinkovics for collaboration at an early stage.
The project was supported by  the Doctoral Excellence Fellowship Programme of NKFIH and its research Grant K134946.
   
\bibliography{BBK-v2}

\end{document}